\magnification=\magstep1 
\baselineskip=14 pt
\hsize=5 in
\vsize=7.3 in
\pageno=1

\vskip 1,8 cm
\centerline{\bf TEMPERATURE NOTION IN A CURVED SPACETIME}
\vskip 1,8 cm 
\centerline{\bf Carlos E. Laciana}
\centerline{\sl Instituto de Astronom\'{\i}a y F\'{\i}sica del Espacio} 
\centerline{\sl Casilla de Correo 67 - Sucursal 28, 1428 Buenos Aires, 
Argentina}
\centerline{\sl E-mail: laciana@iafe.uba.ar}
\vskip 1,8 cm
{\bf Abstract}
\vskip 0,2cm
 Scalar radiation, represented by a massless scalar field in a 
Robertson-Walker metric, is taken into account. By using a weak non minimum 
vacuum definition, the radiation temperature as a time dependent function is 
obtained. When the universe  
evolution is 
nearly but non equal to $t^{1/2}$, it is possible  
to fit the temperature of the  
microwave background. A particular massive 
case is compared with the massless one. When the mass of the matter field 
is next to the Planck one and the time is going to infinite, a similar result 
to the Hawking radiation of the blackhole is obtained.      
\vskip 2,5 cm
\vfill\eject
I.{\bf Introduction}

 The problem of the vacuum definition in Quantum 
Field Theory in Curved Spacetime is discussed 
since more than two decades ago [1].  The non trivial Riemanian 
connection, which appears in the field equation, under changes of the Cauchy
surfaces, 
avoid the invariance, 
of the 
expansion in  normal  modes  of  the  solution. The 
modification of the basis of solutions is represented by the 
Bogoliubov transformations, which are related with the particle creation 
between the space like Cauchy surfaces ``in" and ``out". The conditions  
defining the vacuum  in both surfaces were studied, for example, in refs.  [2],
[3] and [4], where some of the properties of the flat vacuum were generalized 
to curve geometries. The properties of the vacua are formulated via the 
Feynman propagator.  However  there are not cases where  a  time  
dependent definition is
given.  In the present  work  we  will return to the pioneer article of Parker
[5], where a time depending formalism of the Bogoliubov transformation is 
developed. Then we will consider Bogoliubov transformations driven 
by the field equation. The condition that defines the vacuum  
on the ``out" surface will be the 
minimal  vacuum  definition,  used  in  ref.[3],  but  without  imposing  the
minimization of the energy.   That condition is 
incompatible with the dynamical particle creation, as it will be proved.   
Moreover  we  will  see
that a Planckian distribution emerges from the field 
equation without additional conditions.    
\vskip 1,2 cm
   
II. {\bf Field equation and Bogoliubov transformation}
 Let us consider the action for a massive scalar field with arbitrary 
coupling given by  
$$S={1\over 2}\int {\sqrt {-g}} {d^4}x 
({{{\partial}_{\mu}}\varphi}{{{\partial}^{\mu}}{\varphi}} -
({m^2}+\xi R){{\varphi}^2})\eqno (1)$$
\vskip 0,2 cm
in a spacetime described by  
$${ds^2} = {dt^2} - {a(t)^2}({dx^2} + {dy^2} + {dz^2})\eqno (2)$$ 
\vskip 0,2 cm
Therefore the variation of the action results in the field equation 
$$({{\bigtriangledown}_{\mu} {\partial}^{\mu}}  +  {m^2}){\varphi}  =  0\eqno
(3)$$
\vskip 0,2 cm
(with $\mu = 0,1,2,3$). 
\vskip 0,2 cm
 One can expand $\varphi$ using a Fourier representation.  A discretization,  
in the same form as in  
ref.[5], can be used, so we introduce the periodic boundary condition 
$\varphi ({\bf x} + {\bf n} L, t) = \varphi ({\bf x}, t)$, where $\bf n$ is 
a vector with integer Cartesian components and $L$ a length which goes to 
infinity at the end of the calculation.   Then, given a basis of solutions of
the field eq.(3) $\{\psi_{\bf k}(x)\}\bigcup 
\{\!{\psi_{\bf k}}^{\ast}(x)\}$, one can 
expand $\varphi$ in the form  
$$\hat\varphi ({\bf x},t)={\sum_{\bf k}}[A_{\bf k} {\psi}_{\bf k}(x)+ 
{A^{\dagger}}_{\bf k}{\psi^{*}}_{\bf k}(x)]\eqno (4)$$
\vskip 0,2 cm
where $A_{\bf k}$ and ${A^\dagger}_{\bf k}$ are annihilation-creation 
operators which satisfy the bosonic commutation relations
$$[A_{\bf k},A_{{\bf k}^{\prime}}]=0,\ \ 
[{{A^{\dagger}}_{\bf k}},{{A^{\dagger}}_{{\bf k}^{\prime}}}]=0,\ \ 
[A_{\bf k},{A^{\dagger}}_{{\bf k}^{\prime}}]=\delta_{{\bf    k},{\bf
k}^{\prime}}\eqno (5)$$
\vskip 0,2cm
Moreover these operators act on the $|0>$ vacuum in the way  
$$A_{\bf k}|0>=0, \ \ \ {A^{\dagger}}_{\bf k}|0>=|1_{\bf k}>, etc.\eqno (6)$$ 
\vskip 0,2cm

 Due to  the  symmetry  of  the  spacetime,  it  is  possible  to separate the
dependence in the spatial coordinate ${\bf x}$ and the temporal one; then 
we write 
$$\psi_{\bf k}(x)=h_{\bf k}(t)\exp {i{\bf k}.{\bf x}}\eqno (7)$$
\vskip 0,2cm
replacing the last equation in eq.(3), $h_{\bf k}(t)$ satisfy 
$${{\ddot h}_{\bf k}}(t)+{{Q_{\bf k}}^2}(t)h_{\bf k}(t)=0\eqno (8)$$
\vskip 0,2cm
$${Q_{\bf k}}^2(t):=\omega^2-{9\over 4}H^2-{3\over2}{\dot H}+\xi R\eqno (9)$$
\vskip 0,2cm
in eq.(9) $R$ is the curvature scalar, given in the R-W metric as 
$R=6(2H^2+{\dot H})$, with $H={\dot a}/a$ the Hubble coefficient and 
$\omega^2={k^2/a^2}+m^2$. 
 
 Following Parker [5] we introduce the time depending operators 
$a_{\bf k}(t)$ and ${a_{\bf k}}^{\dagger}(t)$, which satisfy eq.(5) for 
for any time, with the boundary condition 
$$A_{\bf k}:=a_{\bf k}(t=t_{0})\eqno (10)$$
\vskip 0,2cm
and operate on the time depending Fock space, as follows:
$$a_{\bf k}(t)|0,t>=0, \ \ \  {a_{\bf k}}^{\dagger}(t)|0,t>=|1_{\bf k},t>, \
\ \ etc.\eqno (11)$$
\vskip 0,2cm

 Now the $\hat\varphi$ is expanded using the time depending operators, i.e.: 
$${\hat\varphi ({\bf x}, t)} = {\sum _{\bf k}}[{a_{\bf k}}(t) 
{\phi _{\bf k}}(x) + {{a^{\dagger}}_{\bf k}}(t) 
{{\! \phi ^{\ast}}_{\bf k}}(x)]\eqno (12)$$ 
\vskip 0,2 cm 
with $\phi$ as a generalization of the normal modes of flat spacetime: 
\vskip 0,2cm
$${{\phi _{\bf k}}(x)} = {1\over {(L a(t))^{3/2} {\sqrt {2W}}}} \exp {i(
{{\bf k}{\bf x}} - {\int _{t_{0}}}^{t} W(k, {t^{\prime}}) dt^{\prime})}
\eqno (13)$$ 
\vskip 0,2 cm

 Replaying the ansatz used in ref.[5] we introduce the complex c-number 
functions of ${\bf k}$ and $t$; $\alpha({\bf k},t)$ and 
$\beta({\bf k},t)$, such that 
$${a_{\bf k}}(t) = {{\alpha ({\bf k},t)}^{\ast}}A_{\bf k}  + 
\beta ({\bf k},t) {A^{\dagger}}_{-{\bf k}}\eqno (14)$$
\vskip 0,2 cm
Then the particle creation operator 
${a^{\dagger}}_{\bf k}(t)a_{\bf k}(t)$ satisfies 
$$n_{\bf k}:=|\beta_{\bf k}|^{2}=
<0|{a^{\dagger}}_{\bf k}(t){a_{\bf  k}(t)}|0>\eqno (15)$$
\vskip 0,2cm
where $n$ is the mean value of the created particles. 

 Following with  a brief review of the ref.  [5], which is necessary 
to understand the main results,  we replace  eq.(14)  in  eq.(12) and by 
comparison with eqs (4) and (7) we obtain  
\vskip 0,2cm
$$h({\bf k},t)={1\over {{{(La(t))}^{3/2}}(2W({\bf k},t))^{1/2}}}
[{\alpha({\bf k},t)}^{*}{e^{-i\int_{t_{0}}^{t}Wd{t^{\prime}}}} + 
{\beta({\bf k},t)}^{*}{e^{i\int _{t_{0}}^{t}Wd{t^{\prime}}}}]\eqno (16)$$
\vskip 0,2cm

 Because that the functions $\psi$ and $\psi^{\ast}$, defined by 
eq.(7), form a basis of solutions, 
then the Bogoliubov coefficients $\alpha$ and 
$\beta$, must satisfy
$$|\alpha|^2-|\beta|^2=1\eqno (17)$$ 
\vskip 0,2cm
Eq. (17) is equivalent to the parametrization:

$$\alpha ({\bf k},t)  =  {e^{-i\gamma_{\alpha}({\bf k},t)}}\cosh \theta ({\bf
k},t)\eqno (18 a)$$
\vskip 0,2 cm
$$\beta ({\bf k},t) =  {e^{i\gamma_{\beta}({\bf  k},t)}}\sinh  \theta  ({\bf
k},t)\eqno (18 b)$$
\vskip 0,2 cm
where $\gamma_{\alpha}$ and $\gamma_{\beta}$ are arbitrary functions, 
and $\theta$ (the Bogoliubov angle) is related with the particle creation 
number by 
$$n_{\bf k}(t)=\sinh^{2}\theta({\bf k},t)\eqno (19)$$
\vskip 0,2 cm

 Putting eqs  (16)  and  (18)  in eq.  (8), performing the separation between
real and imaginary parts, we have the system of equations: 
$$\cosh\theta[(1 + {\tanh \theta}\cos \Gamma)M^{2} + 
2W{{\dot \gamma}_{\alpha}}] = 0\eqno (20a)$$
\vskip 0,2 cm
$$\sinh\theta[{M^{2}}{\sin \Gamma} + 2W{\dot \theta}]= 0\eqno (20 b)$$
\vskip 0,2cm
where
\vskip 0,2cm
$${M^2}:= -{1\over 2}{({\dot W}/W\dot)} + {1\over 4}{({\dot W}/W)^{2}}  
 - {W^{2}}+{Q^{2}}\eqno (21)$$
\vskip 0,2cm
and
\vskip 0,2cm
$$\Gamma:= {\gamma_{\alpha}} + {\gamma_{\beta}} - 
2\int_{t_{0}}^{t}Wdt^{\prime}\eqno (22)$$
\vskip 0,2cm

 In particular eqs(20) are satisfied when 
$$\sinh\theta=0$$
$$\cosh\theta=1$$
\vskip 0,2cm
the first equation corresponds to non-creation of particles, 
which is consistent 
with eq.(20a), because in that case it results
$$M^{2}+2W{{\dot \gamma}_{\alpha}}=0$$
\vskip 0,2cm
but from eq.(18a) it is ${\gamma_\alpha}=0$, then ${M^{2}}=0$. 
In any other case we have the system of equations:  
\vfill\eject
$$(1 + {\tanh \theta}\cos \Gamma){M^{2}} + 2W{{\dot \gamma}_{\alpha}} 
= 0\eqno (23a)$$
\vskip 0,2 cm
$${M^{2}}{\sin \Gamma} + 2W{\dot \theta}= 0\eqno (23b)$$
\vskip 0,2cm
as in this case is $M^{2}\not=0$, we can rewrite eq.(23a) in the form 
$$\tanh\theta\cos\Gamma=-(1+
2{W\over {M^{2}}}{{\dot\gamma}_{\alpha}})\eqno (24)$$
\vskip 0,2cm
from eqs (19) and (17) we can write
$$\tanh\theta=\left(n\over{n+1}\right)^{1/2}$$
\vskip 0,2cm
By replacing the last equation in eq. (24) and putting at square both 
members of that equation, we get 
$${n\over {n+1}}{{\cos}^{2}}\Gamma= (1+
2{W\over {M^{2}}}{{\dot\gamma}_{\alpha}})^{2}\eqno (25)$$
\vskip 0,2cm
Therefore from eq.(25) it is obtained  
$$n={1\over{f-1}}\eqno (26)$$
\vskip 0,2cm
with
\vskip 0,2cm
$$f={{{\cos^{2}}\Gamma}\over{(1+
2{W\over {M^{2}}}{{\dot\gamma}_{\alpha}})^{2}}}\eqno (27)$$
\vskip 0,2cm
As we can see from eq.(27) 
$f=f({\gamma_\alpha},{{\dot\gamma}_\alpha},{\gamma_\beta},W,{\dot W})$, 
then we have superabundance of variables, we can choose some of them in a 
convenient way. 
 If we use unities such that $c={k_B}=\hbar=1$, we can define temperature 
by the equalization of the function $f$ to the Planckian functional form;
 
$$f=\exp ( {\epsilon_{k}}/T)\eqno (28)$$
\vskip 0,2cm 
with $\epsilon_{k}$  the  energy  by  mode, $T$ the temperature.
From eq. (28) we can calculate  the temperature $T$ as a function of the 
particle model used, as we will see in the next section.    
 
\vfill\eject 
III. {\bf Time depending temperature}
\smallskip
  To get an expression for the temperature, first we must  
calculate the energy by  mode  $\epsilon_{k}$.  That can be obtained from the
total energy, i.e. 
$$E=<0|{\hat H}|0>\eqno (29)$$
\vskip 0,2cm
${\hat H}$ is the metric hamiltonian, which can be defined by means of the 
$00$ component of the energy-momentum operator ${\hat T}_{\mu \nu}$, which is 
obtained by the variation of the action respect to the metric [1]: 
$${\hat H}=\int {a^3}{d^3}x{\hat T}_{00}\eqno (30)$$
\vskip 0,2cm
${\hat T}_{00}$ is a functional of the field operator $\hat \varphi$. 
In the discretized formulation is $\int {d^3}x=L^3$.  Following ref.[6] we 
can drop the oscillatory term  in the $k$ variable, obtaining an expression
for the energy similar to the one corresponding to a set of independent 
quantum oscillators:
$$E=\sum_{\bf k}({1\over 2}+n_{\bf k}){\epsilon_{\bf k}}\eqno (31)$$
with (see [4]);
$${{\epsilon}_{\bf k}}={1\over {2W}}\{{1\over 4}{[{{\dot W}\over W}
+3(1-4\xi)H]^2}+6\xi{H^{2}}(1-6\xi)+
{W^2}+{\omega^2}\}\eqno (32)$$
\vskip 0,2cm

 Then from  eq.(28)  we  can  look  for a temperature as a function of the
vacuum definition (or the particle model), in the form 
$$T={{\epsilon_{k}}\over {2\ln [\cos \Gamma/(1+
2W{\dot\gamma}/{M^{2}})]}}\eqno (33)$$
\vskip 0,2cm

 In order to have well defined the logarithm, the argument must be bigger 
than zero. Using the arbitrariness in the phase $\Gamma$, we can choose 
$$\Gamma\equiv \pi/4 \eqno (34)$$
\vskip 0,2cm
 
 Then the field equations turn to be 
\vfill\eject 
$$(1 + {1\over \sqrt {2}}{\tanh \theta}){M^{2}} + 2W{{\dot \gamma}_{\alpha}} 
= 0\eqno (35a)$$
\vskip 0,2 cm
$${1\over \sqrt {2}}{M^{2}} + 2W{\dot \theta}= 0\eqno (35b)$$
\vskip 0,2cm

 By the integration of eq.(35b) and by the replacing in eq.(35a) we 
obtain
$${{\dot\gamma}_\alpha}={-{M^{2}}\over {2W}}[1+
{1\over {\sqrt{2}}}\tanh\left(B-
{1\over{2\sqrt{2}}}\int_{t_{0}}^{t}{{M^{2}}\over W}dt\right)]\eqno (36)$$
\vskip 0,2cm
B is an integration constant, which is related with the particles 
that are present 
at the time $t=t_{0}$, i.e. 
$$n(t=t_{0})={\sinh^{2}}B$$
\vskip 0,2cm
Replacing eq.(36) in eq.(33), we have 
$$T={{\epsilon_{k}}\over {2\ln[\coth[u]]}}\eqno (37)$$
\vskip 0,2cm
with
\vskip 0,2cm
$$u:={1\over {2\sqrt{2}}}{\int_{t_{0}}^{t}}{{M^{2}}
\over {W}}dt^{\prime}-B\eqno (38)$$
\vskip 0,2cm
where ${\epsilon_{k}}={\epsilon_{k}}[W]$ is given by eq.(32) and 
$M^{2}=M^{2}[W]$ is given by eq.(21). In order to introduce 
the particle model we 
must to give the function $W$. In our case we will propose 
$$W=\omega\eqno (39)$$
\vskip 0,2cm

 It is interesting to note that eq.(37) is very similar to the one obtained 
in a parametric photon pair production [7]. That photons has the same 
statistical that our field modes interacting with the curved geometry.

 Eq.(39) means that at the observation time  $t$, the function $W$ 
satisfies the zero WKB order [2],[3]. Then the modes $\phi$ are analogous 
to the ones of flat spacetime. This is motivated by the fact that 
at the present time  
the universe is very flat. Then if we look at eq.(12) we have formally an  
expression for the field operator similar than the one corresponding to 
scalar bosonic field in flat space time. We cannot use the minimization 
of energy criterium, as we will see latter.    
That criterium is only compatible with an in-out process,   
but not when the dynamics is considered.
\bigskip
IV. {\bf Calculation of T in some particular cases}
\smallskip

 We will now study some particular cases in order to test the physical 
behavior of eq.(37). 
\item { i)} Firstly let us considered the massless minimally coupled case 
$(m=0,\xi=0)$. Then from eq.(39) we have 
$$W=k/a\eqno (40)$$
\vskip 0,2cm
We also suppose that at the beginning there are not particles, 
then $B=0$ in eq.(38). Performing the calculation it results 
$${\epsilon_{k}}={a\over {2k}}({H^2}+2k^2/a^2)\eqno (41)$$
\vskip 0,2cm
$$u=-{1\over {2\sqrt{2}}}{{\int^{t}}_{t_{0}}}a(2{H^2}+
{\dot H})dt^{\prime}\eqno (42)$$
\vskip 0,2cm
For the universe evolution given by 
$$a(t)=a_{0}\left(t\over {t_{0}}\right)^{\alpha}\eqno (43)$$
\vskip 0,2cm
then 
$$\epsilon_{k}={k\over a_{0}}\left(
{{{a^{2}_{0}}{\alpha^{2}}}\over {2k^2{t_{0}}^2}}x^{2-\alpha}
+{x^\alpha}\right)\eqno (44)$$
\vskip 0,2cm
$$u={1\over {2\sqrt{2}}}{{a_{0}\alpha}\over {k t_{0}}}
{{(2\alpha -1)}\over {(\alpha-1)}}(1-x^{1-\alpha})\eqno (45)$$
\vskip 0,2cm
where $x:=t_{0}/t$. 

 From eqs (45) and (37) it results that $T$ can be well defined when 
$\alpha\in[0,1/2)$.

 The semiclassical approach is valid until the Planck time [1], 
therefore if we choose $t_{0}$ as the Planck time the variable $x\in[0,1]$.

  From fig.1  we can see that if there are not particles at the initial time, 
the temperature begins from zero and increases to reach a maximum, after that
 goes to zero (in the massless case). The qualitative behavior is similar in
all the interval of $\alpha$, in the fig. 1 we compare the cases 
$\alpha=1/4$  (dot-dashed  line)    and   $\alpha=0.49$  (full  line).    The
calculation was simplified by using the relation
$$t_{0}{k\over a_{0}}\sim 1\eqno (46)$$
\vskip 0,2cm
We are thinking the universe, at the Planck time, as analogous to a radioactive 
nucleus with $t_{0}$ as the mean time life and 
$\Delta \epsilon_{k}\sim k/{a_{0}}$ the band width  of the energy [8].  That
approach means that it is necessary to wait a time $t_{0}$ in 
order to generate a particle of energy $k/{a_{0}}$.   
\vskip 0,2 cm
\item{ ii)} Now we consider the massive minimally coupled case ($m\not=0$, 
$\xi=0$). Then instead of eq. (45) we have 
$$u={{{a_{0}}\alpha (1-2\alpha)}\over {2\sqrt{2}k{t_{0}}}}(I_{1}(1)-
I_{1}(x))
+{{{m^2}\alpha(1-2\alpha)}\over {4\sqrt{2}{k^{3}}{t_{0}}}}(I_{2}(1)-
I_{2}(x))$$ 
$$+ {{3{m^4}{{a_{0}}^{5}}\alpha}\over {8\sqrt{2}
{t_{0}}{k^5}}}(I_{3}(1)-I_{3}(x))\eqno (47)$$
\vskip 0,2 cm
where
$$I_{1}(x)=\int^{x}{{x^{-\alpha}}\over {(1+\lambda x^{-2\alpha})^{1/2}}}dx,
\ \ I_{2}(x)=\int^{x}{{x^{-3\alpha}}
\over {(1+\lambda x^{-2\alpha})^{3/2}}}dx,$$
$$I_{3}(x)=\int^{x}{{x^{-5\alpha}}
\over {(1+\lambda x^{-2\alpha})^{5/2}}}dx,$$
\vskip 0,2 cm
with $\lambda:= m^2{a_{0}}^2/k^2$ an adimensional 
parameter. In order to compare with the 
massless case we can do the calculation for the particular case 
$\alpha=1/4$.  Using  $\lambda\sim  1$ (which corresponds to $m\simeq m_{p}$)
to consider an extremal case. The result is also shown in fig. 1 
(dashed line).  It  is  interesting  to  note  that  when  the  time  goes to
infinity, in the massive case, the temperature goes to a constant approximately 
equal to $m_{p}/4$. 
That result is analogous to the Hawking one of the black 
hole with mass $m_{p}$ (except that the $2\pi$ factor not appears), remember 
that $T_{H}=1/{8\pi GM}$ [9]. 
\bigskip
V.{\bf Conclusions and comments}
\smallskip

 The main result is that from the field equation and with the more 
straightforward generalization of the definition, for the modes in  
Quantum Field Theory in flat spacetime, we can get a time depending 
temperature  expression.    It  is    really    the    more  straightforward
generalization  because  we  stress  for  the    observation   time  $t$  the
correspondence 
$$W\to \omega$$
\vskip 0,2 cm
where $\omega$ for $t=cte$ corresponds to the frequency of the field mode 
in the Minkowski space. 

 As we can see from fig.1, the  temporal  behavior  of the $T$ function, for
the scalar radiation, is physically reasonable, and lead us to fit the 
microwave background by means of the fine tuning of the $\alpha$ parameter, 
which must be very close but not equal to 1/2, as it results from eq.(45). 
 
 For the massive case the temperature goes to  a constant value when the time
increase to infinity (see the dashed line in fig. 1).     
 
 Let me do now a brief comment about the minimization energy conditions as a 
time depending constraint. For an arbitrary coupling the minimization energy 
condition (or hamiltonian diagonalization condition) [4] is: 
$${W^{2}}=6\xi(1-6\xi)H^{2}+{\omega^{2}}\eqno (48a)$$
$${{\dot W}\over W}=-3(1-4\xi)H\eqno (48b)$$
\vskip 0,2 cm
this equations, for all the time, in the massive case, has not physical sense 
because  the  scalar factor of the universe must depend on the
mass and the  mode,  i.e.  ;  $a=a(k,m,t)$.  Still in the massless case it has
not physical meaning when  $\xi\not=0,  1/6$.    If  $\xi=0$ and $m=0$, eq.
(48b) $\Rightarrow a(t)= constant$.  Only when $\xi=1/6$ (conformal coupling) 
eq. (48b) is satisfied without any condition on $a(t)$. But if we  
consider eq. (21), from the conditions given by eqs (48) it results 
$${M^{2}}=-12\xi{H^{2}}(1-6\xi)\eqno (49)$$
\vskip 0,2cm
Then when $\xi=1/6$ we have ${M^{2}}=0$, therefore from eq.(20b) we have 
$${\dot\theta}=0\eqno (50)$$
\vskip 0,2cm
So from eq.  (19) we can say that in this case the particle creation is null. 
 Then when the dynamics is considered, the criterium given by eqs. (48) 
does not work.    
      
\bigskip
{\bf ACKNOWLEDGEMENTS} 

 This work was partially supported by the Concejo Nacional de Investigaciones 
Cient\'{\i}ficas y T\'ecnicas (Argentina) and   
by  the  European  Community  DGXII. 
\bigskip

{\bf REFERENCES}
\vskip 0,2cm
   
\item{$[1]$} N.D.Birrell and P.C.W.Davies;  ``Quantum Fields in Curved Space" 
(Cambridge University Press, Cambridge, England, 1982).
\item{$[2]$}  E.Calzetta  and  M.Castagnino, Phys.  Rev.  D  {\bf  28},  1298
(1984).
\item{$[3]$} M.Castagnino and F.D.Mazzitelli, Phys. Rev. D {\bf 31}, No 4, 
742, (1985).  
\item{$[4]$} M.Castagnino  and  C.Laciana;    J.  Math.  Phys.  {\bf 29} (2),
(1988), pp 460-475.
\item{$[5]$} L.Parker;  Phys.Rev., Vol.183, $N^{o}$ 5, (1969), pp 1057-1068.  
\item{$[6]$} C.E.Laciana; Gen. Rel. and Grav., Vol. 28, No. 8, (1996), 
pp 1013-1026.
\item{$[7]$} B.Yurke and M.Potasek, Phys. Rev. A {\bf 36}, 3464 (1987).    
\item{$[8]$} A.Messiah, ``M\'ecanique quantique" Tome I (Dunod, Paris, 1969), 
page 117.    
\item{$[9]$} J.V.Narlikar and T.Padmanabhan, Gravity, Gauge Theories, and 
Quantum Cosmology (Reidel, 1986).  
  
\vfill\eject
\centerline{\bf Figure Captions}
\bigskip
\item { FIG. 1} Dimensionless temperature $T/b$, with $b= k/2a_{0}$, vs 
the parameter $x=t_{0}/t$. The full line corresponds to the 
massless case with $\alpha=0.49$, the dashed line to the massless with 
$\alpha=1/4$ and the dot-dashed line to the massive case with $\alpha=1/4$. 
 
\bye